\documentclass[twocolumn,showpacs,preprintnumbers,amsmath,amssymb]{revtex4}

\usepackage{graphicx}
\usepackage{dcolumn}
\usepackage{bm}

\begin{document}


\title{Clustering in light nuclei in fragmentation above 1 A GeV}

\author{N.~P.~Andreeva}
 \affiliation{Institute for Physics and Technology, Almaty, Republic of Kazakhstan}
\author{D.~A.~Artemenkov}
   \affiliation{Joint Insitute for Nuclear Research, Dubna, Russia}
 \author{V.~Bradnova}
   \affiliation{Joint Insitute for Nuclear Research, Dubna, Russia} 
\author{M.~M.~Chernyavsky}
  \affiliation{Lebedev Institute of Physics, Russian Academy of Sciences, Moscow, Russia} 
 \author{A.~Sh.~Gaitinov}
   \affiliation{Institute for Physics and Technology, Almaty, Republic of Kazakhstan}
\author{N.~A.~Kachalova}
   \affiliation{Joint Insitute for Nuclear Research, Dubna, Russia} 
\author{S.~P.~Kharlamov}
   \affiliation{Lebedev Institute of Physics, Russian Academy of Sciences, Moscow, Russia}
\author{A.~D.~Kovalenko}
   \affiliation{Joint Insitute for Nuclear Research, Dubna, Russia}  
\author{M.~Haiduc}
   \affiliation{Institute of Space Sciences, Magurele, Romania}
 \author{S.~G.~Gerasimov}
   \affiliation{Lebedev Institute of Physics, Russian Academy of Sciences, Moscow, Russia} 
\author{L.~A.~Goncharova}
   \affiliation{Lebedev Institute of Physics, Russian Academy of Sciences, Moscow, Russia} 
\author{V.~G.~Larionova$^{\dag}$} 
 \affiliation {Lebedev Institute of Physics, Russian Academy of Sciences, Moscow, Russia}      
\author{A.~I.~Malakhov}
   \affiliation{Joint Insitute for Nuclear Research, Dubna, Russia} 
\author{A.~A.~Moiseenko}
   \affiliation{Yerevan Physics Institute, Yerevan, Armenia}
\author{G.~I.~Orlova}
   \affiliation{Lebedev Institute of Physics, Russian Academy of Sciences, Moscow, Russia} 
\author{N.~G.~Peresadko}
   \affiliation{Lebedev Institute of Physics, Russian Academy of Sciences, Moscow, Russia} 
\author{N.~G.~Polukhina}
   \affiliation{Lebedev Institute of Physics, Russian Academy of Sciences, Moscow, Russia} 
\author{P.~A.~Rukoyatkin}
   \affiliation{Joint Insitute for Nuclear Research, Dubna, Russia} 
\author{V.~V.~Rusakova}
   \affiliation{Joint Insitute for Nuclear Research, Dubna, Russia} 
\author{V.~R.~Sarkisyan}
   \affiliation{Yerevan Physics Institute, Yerevan, Armenia} 
\author{T.~V.~Shchedrina}
   \affiliation{Joint Insitute for Nuclear Research, Dubna, Russia}
\author{E.~Stan} 
   \affiliation{Institute of Space Sciences, Magurele, Romania}
\author{R.~Stanoeva}
  \affiliation{Institute for Nuclear Research and Nuclear Energy, Sofia, Bulgaria}
 \author{I.~Tsakov}
   \affiliation{Institute for Nuclear Research and Nuclear Energy, Sofia, Bulgaria}
 \author{S.~Vok\'al}
   \affiliation{P. J. \u Saf\u arik University, Ko\u sice, Slovak Republic}
\author{A.~Vok\'alov\'a}
   \affiliation{P. J. \u Saf\u arik University, Ko\u sice, Slovak Republic}    
 \author{P.~I.~Zarubin}
     \email{zarubin@lhe.jinr.ru}    
     \homepage{http://becquerel.lhe.jinr.ru}
   \affiliation{Joint Insitute for Nuclear Research, Dubna, Russia} 
 \author{I.~G.~Zarubina}
   \affiliation{Joint Insitute for Nuclear Research, Dubna, Russia}   

\date{\today}

\begin{abstract}
\indent The relativistic invariant approach is applied to analyzing the 3.3~A~GeV $^{22}$Ne fragmentation in a nuclear track emulsion.
 New results on few-body dissociations have been obtained from the emulsion exposures to 2.1~A~GeV $^{14}$N and  1.2~A~GeV $^9$Be nuclei.
 It can be asserted that the use of the invariant approach is an effective means of obtaining
 conclusions about the behavior of systems involving a few He nuclei at a relative energy close 
 to 1~MeV per nucleon. 
The first observations of fragmentation of 1.2~A~GeV $^8$B and $^9$C nuclei in emulsion are described.
 The presented results allow one to justify the development of few-body aspects of nuclear astrophysics.

\end{abstract}

\maketitle
\section{\label{sec:level1}Introduction}
\indent Interactions in few-body systems consisting of more than two $^{1,2}$H and $^{3,4}$He nuclei can contribute to 
 a nucleosynthesis pattern.
 A macroscopic medium composed of the lightest nuclei having energy of a nucleosynthesis scale
 can possess the properties analogous to those of 
dilute quantum gases of atomic physics. In this sense, few-body fusions imply a phase transition 
to \lq\lq drops\rq\rq  of a quantum liquid, that is, to heavier nuclei. 
 Fusions can proceed via the states corresponding to low-lying cluster excitations in forming nuclei. \par
 \indent On a microscopic level the phase transition can proceed through the production of quasi-stable and
 loosely bound quantum states.
 Among candidates for such states one can consider the dilute $\alpha$ particle Bose condensate
 \cite{Horiuchi03} as well as radioactive and
 unbound nuclei along a proton drip line. 
At first glance, exploration of few-body transitions in the laboratory conditions seems to be impossible.
 Nevertheless, such processes can indirectly be explored in the inverse processes of relativistic 
nucleus breakups in a nuclear track emulsion by selecting the excitations close to the few-body decay threshold. \par
 \indent Experimental data on 
charged topology for the final states of a number of light nuclei have been described 
 in \cite{El-Naghy88,Baroni90,Baroni92,Jilany04,Andreeva05}.
 More specific studies were performed for the leading channels like
$^{12}$C$\rightarrow$3$\alpha$ \cite{Belaga95}, 
 $^{16}$O$\rightarrow$4$\alpha$ \cite{Andreeva96,Glagolev04}, 
$^{6}$Li$\rightarrow$d$\alpha$ \cite{Lepekhin98,Adamovich99},
$^{7}$Li$\rightarrow$t$\alpha$ \cite{AdamovichPh04},
$^{10}$B$\rightarrow$d$\alpha\alpha$ \cite{Adamovich04}, 
and $^{7}$Be$\rightarrow^{3}$He$\alpha$ \cite{BradnovaA04}. A collection of appropriate reaction 
images can be found in  
\cite{Bradnova04,web}. \par
  \begin{figure*}
    \includegraphics{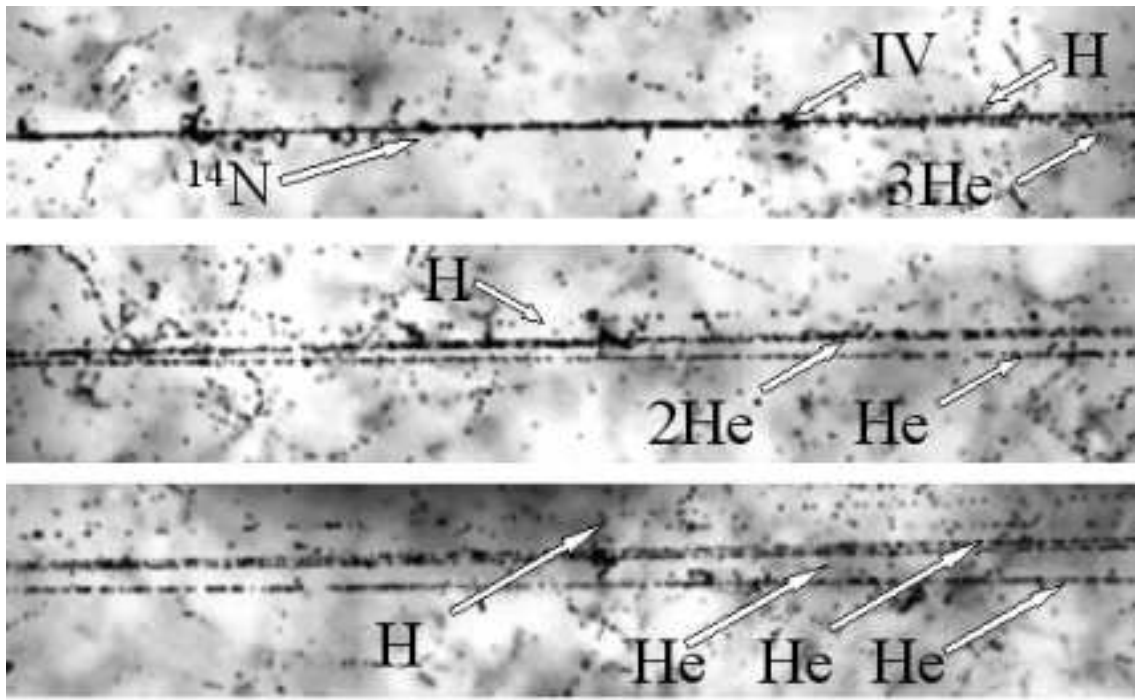}
    \caption{\label{fig:1}Example of peripheral interaction of a 2.1~A~GeV $^{14}$N nucleus in a nuclear track emulsion (\lq\lq white\rq\rq ~star).
     The interaction vertex (indicated as {\bf IV}) and nuclear fragment tracks ({\bf H} and {\bf He}) in a narrow angular
     cone are seen  on the upper microphotograph.
     Following the direction of the fragment jet, it is possible to distinguish 1 singly and 3 doubly charged 
     fragments on the middle and bottom microphotograph.}
    \end{figure*}
 \indent In the present paper the behaviour of relativistic systems consisting of several H and He nuclei will be described
 in terms of invariant variables of a 4-velocity space as suggested in \cite{Baldin90}. The invariant presentation makes it possible
 to extract qualitatively new information about few-cluster
 systems from fragmentation of relativistic nuclei in peripheral interactions. 
The invariant approach is applied to the existing data on 3.3~A~GeV $^{22}$Ne interactions in
 a nuclear track emulsion,
 as well as to new data for 2.1~A~GeV $^{14}$N and 1.2~A~GeV $^{9}$Be nuclei extracted from a portion of
 recently exposured emulsion.
 The first observations of the fragment topology for neutron-deficient $^{8}$B and $^{9}$C nuclei in
 emulsion are described
 in this report. Last emulsion exposures were performed at the JINR Nuclotron in the years 2002-4
 \cite{Malakhov04}.\par

\section{\label{sec:level2}Nuclear fragment jets}

 \indent The relativistic projectile fragmentation results in the production of a
 fragment jet which can be defined by invariant variables characterizing relative motion

\begin{equation}
{b_{ik}=-\left( {P_i \over m_i} -{P_k \over m_k}\right)^2}
\end{equation}
 \newline with $P_{i(k)}$ and $m_{i(k)}$  being
 the 4-momenta and the masses of the $i$ or $k$ fragments.
 Following \cite{Baldin90}, one can suggest that a jet is composed of the nuclear fragments having relative motion within the non-relativistic range
 10$^{-4}$~$<$b$_{ik}<$~10$^{-2}$.
 The lower limit corresponds to the ground state decay $^{8}$Be$\rightarrow$2$\alpha$, while the
 upper one - to the boundary of low-energy nuclear interactions. The expression of the data via the
 relativistic invariant variable b$_{ik}$ makes it
 possible to compare the target and projectile fragmentation in a common form.
 Fig.~\ref{fig:1} shows the microphotograph of a special example of a projectile fragment jet - the \lq\lq white\rq\rq star 
as introduced in \cite{Baroni90}.
 It corresponds to the case of a relativistic nucleus dissociation accompanied by neither a target fragment
 nor meson production.\par
 \indent The variable characterizing the excitation of a fragment jet as a whole is an invariant mass
 $M^*$ defined as 
 \begin{equation}
   {M^{*2}=(\Sigma P_j)^2=\Sigma (P_i\cdot P_k)}
 \end{equation}
 \newline The system excitation can be characterized also by
 \begin{equation} 
   {Q=M^{*}-M}
 \end{equation}  
 \newline with $M$ being the mass of the ground state of the nucleus
 corresponding to the charge and weight of the fragment system.
 The variable $Q$ corresponds to the excitation energy of the system of fragments in their c. m. s. A useful option
 is 
\begin{equation}    
  {Q\rq={(M^{*}-M\rq)\over A}}
\end{equation}   
   \newline with $M\rq$ being the sum of fragment masses and $A$ the total atomic weight. 
 The normalized variable $Q\rq$ characterizes a mean kinetic energy of fragments per nucleon in their c. m. s.
 Precision of the experimental $b_{ik}$ and $Q$ values  is  
 driven to a decisive degree by the angular resolution in the determination of unit vectors defining the
 direction of the fragment emission. \par
 \indent Due to excellent spatial resolution (about 0.5 $\mu$m)
the emulsion technique is known to be most adequate for the observation and angular measurements of
 projectile fragments down to a total breakup of relativistic nuclei.
 Nevertheless, it has restrictions on the determination of the 4-momentum components of fragments.
 Firstly, the fragment spatial momentum in the projectile fragmentation cone is suggested to be equal
 within a few percent error to the primary
 nucleus value when normalized to the nucleon numbers. Secondly, by 
multiple scattering measurements it is possible to identify the mass  only
 for relativistic H
 isotopes and much harder for He ones. Normally, the $\alpha$
 particle mass is taken for the mass of doubly charged
 fragments in a narrow fragmentation cone. Both assumptions are proven to be reasonable for light stable
 nuclei.\par 

\section{\label{sec:level3}Fragmentation of  $\bf ^{22}$N\lowercase{e}  nuclei}

 \indent  A nuclear state analogous to the dilute Bose gas can be revealed in the formation of $n\alpha$
 particle ensembles possessing quantum coherence near the production threshold.
 The predicted property of these systems is a narrow velocity distribution in the c. m. s.
 \cite{Horiuchi03}.
 Originating from relativistic nuclei, they can appear as narrow $n\alpha$ jets in the forward 
cone defined
 by the nucleonic Fermi motion.
 The determination of the c. m. s. for each event is rather ambiguous while analysis of jets in the
 b$_{ik}$ space enables one to explore $n\alpha$ particle systems in a universal way.\par  

\begin{figure}
\resizebox{0.5\textwidth}{!}{
   \includegraphics{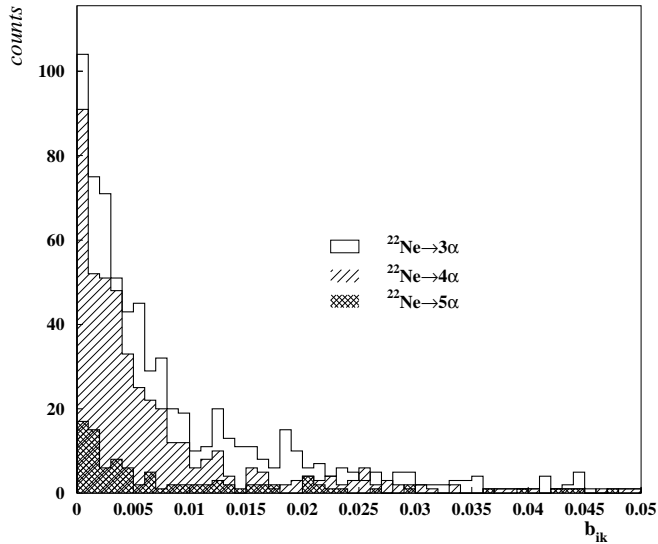}}
      \caption{\label{fig:2}Distribution of $\alpha$ particle pairs $vs$ relative variable 
       $b_{ik}$ (1) for the fragmentation modes $^{22}$Ne$\rightarrow$ $n\alpha$. }
  \end{figure}
 \begin{figure}
\resizebox{0.5\textwidth}{!}
{
  \includegraphics{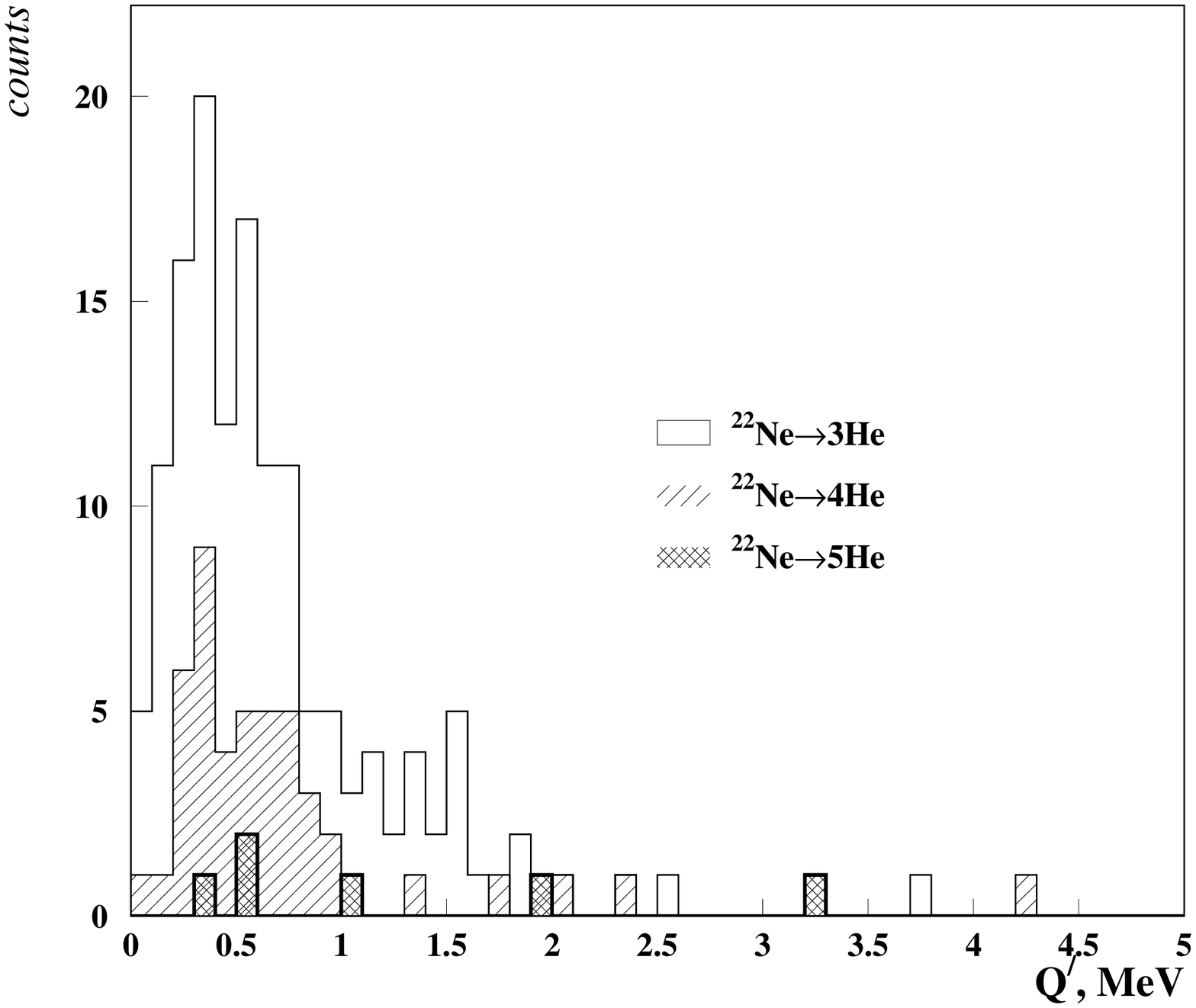}
}
\caption{\label{fig:3}Distribution of $\alpha$ particle pairs $vs$ $Q\rq$ (4) for the fragmentation modes $^{22}$Ne$\rightarrow$ $n\alpha$.}
\end{figure}

\indent At our disposal there are data on 4100 events from 3.3~A~GeV $^{22}$Ne nucleus interactions 
with emulsion nuclei (presented in \cite{El-Naghy88}) which
 contain the classification of secondary tracks by ionization and angles. The 
key feature for the $^{22}$Ne fragmentation consists in a suppression of binary splittings into medium charge fragments 
with respect to He and H cluster formation.
 The increase of a fragmentation degree is revealed 
in a growth of the $\alpha$ particle multiplicity.
 Thus, transition to the
 $n\alpha$ particle states having a high level density predominate over the binary splittings occuring at lower energy thresholds.\par
 \indent In the present analysis, the doubly charged particles found in a forward 6$^{\circ}$ cone were
 classified as relativistic $\alpha$ particles. Fig.~\ref{fig:2} shows the $b_{ik}$ distribution (1) for the
 fragmentation channel $^{22}$Ne$\rightarrow$ $n\alpha$ for n equal to 3 (240 events), 4 (79 events),
 and 5 (10 events) which is rather narrow.
 The distribution \lq\lq tails\rq\rq appear to be due to the $^4$He diffractive scattering or $^3$He formation proceeding at  higher
 momentum transfers. The events, satisfying the non-relativistic criterion $b_{ik}<$10$^{-2}$ for
 each $\alpha$ particle pair were selected for $n$ equal to 3 (141 events), 4 (47 events), and
 5 (6 events).\par
 \indent  Fig.~\ref{fig:3} presents the  normalized $Q\rq$ distribution  (4) for them. Being considered as estimates of the 
 mean kinetic energy per nucleon in the center-of-mass of n$\alpha$ system, the $Q$ values does not exceed
 the Coulomb barrier values.
 Thus, in spite of the high $n\alpha$ multiplicity, the $n\alpha$ jets are seen to remain rather
 \lq\lq cold\rq\rq and consimilar. Among 10 $^{22}$Ne$\rightarrow$5$\alpha$ events  there were found 3 \lq\lq white\rq\rq ~stars. Of these, in 2 \lq\lq golden\rq\rq events $\alpha$ particle tracks are contained even within a 1$^{\circ}$ cone.
 For these two events the value of $Q\rq$ is estimated to be as low as 400 and 600~keV per nucleon. The detection of
 such \lq\lq ultracold\rq\rq ~5$\alpha$ states is a serious argument
 in favor of the reality of the phase transition of $\alpha$ clusterized nuclei to the dilute Bose gas of $\alpha$
 particles. It gives a special motivation to explore lighter n$\alpha$ systems produced
 as potential \lq\lq building blocks\rq\rq of the dilute $\alpha$  particle Bose gas.\par

\section{\label{sec:level4}Fragmentation of $ \bf^{14}$N nuclei}
\indent We are presently engaged in accumulating statistics on the interactions of 2.1~A~GeV $^{14}$N 
nuclei in emulsion impacted on  \lq\lq white\rq\rq
 ~star searches. 
25 \lq\lq white\rq\rq ~stars have already been
 found among 540 inelastic events by  scanning over primary tracks. Such a systematic scanning
 allows one to estimate relative probabilities 
of various fragmentation modes. The secondary tracks of \lq\lq white\rq\rq stars were selected to be concentrated
 in a forward 8$^\circ$ cone.
\begin{table}
\caption{\label{tab:1}Charge-topology distribution of the \lq\lq white\rq\rq ~stars originated from the dissociation 
of 2.1~A~GeV $^{14}$N nuclei.}

\begin{tabular}{l|c|c|c|c|c|c|c|c}
\hline\noalign{\smallskip}
\hline\noalign{\smallskip}
$Z_f$ &6 &5 &5 &4 &3 &3 &-- &-- \\
$N_{1}$ & 1 & -- & 2 & 1 & 4 & 2 & 3 & 1 \\
$N_{2}$ & -- & 1 & -- & 1 & -- & 1 & 2 & 3  \\
$N_{ws}$ & 6 & 2 & 3 & 1 & 1 & 1 & 1 & 10  \\
\hline\noalign{\smallskip}
\hline\noalign{\smallskip}
\end{tabular}
\end{table}
 \par
 \indent Table~\ref{tab:1} shows a number of the found \lq\lq white\rq\rq stars $N_{ws}$ 
composed of a single heavy fragment 
having charge $Z_f$ and  of $N_{1}$ singly and $N_{2}$ doubly charged fragments. The predominant role of the 4-prong mode 
3He+H among  the \lq\lq white\rq\rq ~stars is clearly seen. It 
implies that the exploration  of the 3$\alpha$ systems originated in $^{14}$N fragmentation is rather promising.\par
\begin{figure}
\resizebox{0.5\textwidth}{!}{
   \includegraphics{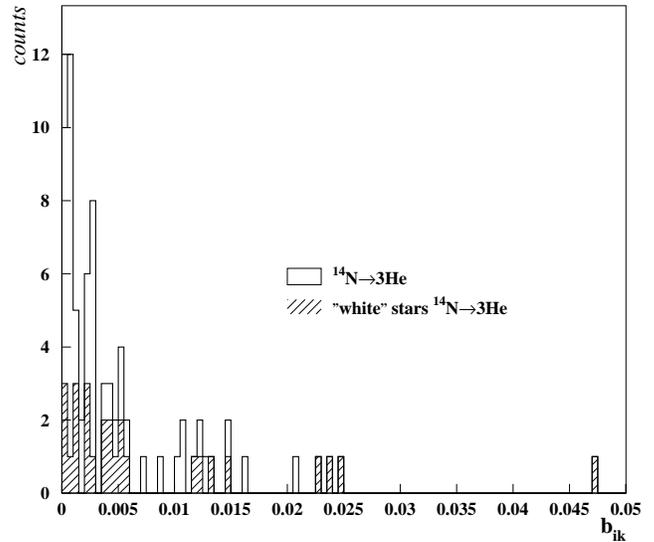}}
      \caption{\label{fig:4}Distribution of $\alpha$ particle pairs $vs$ relative variable 
$b_{ik}$ (1) for the fragmentation mode $^{14}$N$\rightarrow$3$\alpha$.}
  \end{figure}

\indent Fig.~\ref{fig:4}  allows one to compare the $b_{ik}$ distribution for the "white" 3$\alpha$ stars to the case of  3He+H events
where a prohibition on a target fragmentation is lifted of. In both cases, the criterion of a non-relativistic character of
fragment interactions is satisfied.
\par
  
 \begin{figure}
\resizebox{0.5\textwidth}{!}
{
  \includegraphics{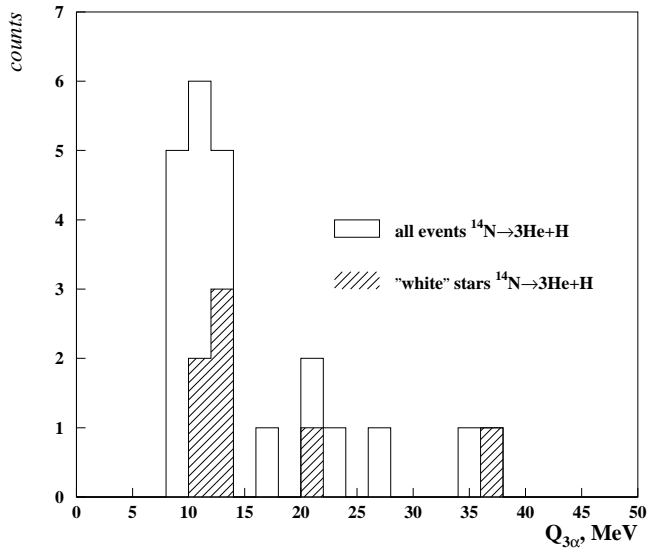}
}
\caption{\label{fig:5}Distribution  of $\alpha$ particle triplets $vs$ $Q_{3 \alpha}$ (3)
for the fragmentation mode $^{14}$N$\rightarrow$3$\alpha$+H.}
\end{figure}
\indent 
 Fig.~\ref{fig:5} shows the $Q_{3\alpha}$ distribution in which the excitation energy  
is counted out from the $^{12}$C nucleus mass. It can be concluded that the major fraction of entries is concentrated within
 a range of 10 to 14~MeV covering the known $^{12}$C nucleus levels. Softening of the selection
 conditions for 3He+H events, under which the target fragment formation is allowable, does not result in a shift of the position of
 the 3$\alpha$ excitation peak.
 This circumstance points out the universality of the mechanism of population of 3$\alpha$ particle
 states.
 Besides, one can readily estimate that the normalized values $Q\rq_{3 \alpha}$ are of the same magnitude as in fig.~\ref{fig:3}.
 Another conclusion is that the contribution of the $\alpha$-$^8$Be state in the 3$\alpha$ configuration
 does not exceed a 10\% -level. This topic awaits for higher statistics allowing a reliable $^8$Be identification. 
\begin{figure}
\resizebox{0.5\textwidth}{!}
{
  \includegraphics{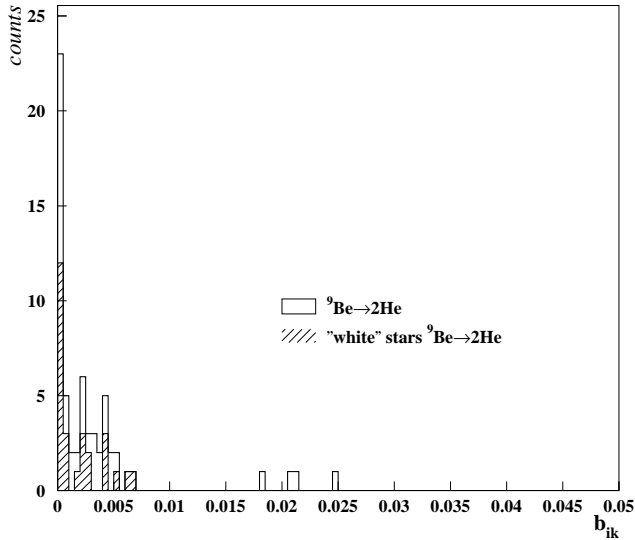}
}
\caption{\label{fig:6}Distribution of $\alpha$ particle pairs $vs$ relative variable 
$b_{ik}$ (1) for the fragmentation mode $^{9}$Be$\rightarrow$2$\alpha$.}
\end{figure}

\section{\label{sec:level5}Fragmentation of $\bf^9$B\lowercase{e} nuclei}
\indent The relativistic $^9$Be nucleus fragmentation is an attractive source for $^8$Be generation due
 to the absence of a combinatorial
 background. The $^8$Be nucleus reveals itself in the formation of $\alpha$ particle pairs having an
 extremely small opening angle of the order of a few 10$^{-3}$ rad in the range of a few GeV.
 The estimation of the $^8$Be production probability will make it possible to clear up the interrelation
 between n-$^8$Be and $\alpha$-n-$\alpha$ excitation modes which
 are important in understanding the $^9$Be structure and fragmentation of heavier nuclei.\par
\indent A secondary $^9$Be beam was produced through fragmentation of the primary 1.2~A~GeV $^{10}$B beam.
 In scanning emulsion layers exposed to $^9$Be nuclei, about 200 interactions are detected with He pair produced
 in a forward
 8$^{\circ}$ cone. As in the previously considered cases, the $b_{ik}$ distribution 
for 50 measured events, which is shown  
in fig.~\ref{fig:6}, confirms the non-relativistic behavior of the relative motion of the produced $\alpha$ particles. 
In just the same way
 as in the case of the $^{14}$N nuclei, softening of the criterion of
 selection of the 2He pairs less rigid does not lead to changes of the distribution shape. \par
\begin{figure}
\resizebox{0.5\textwidth}{!}
{
  \includegraphics{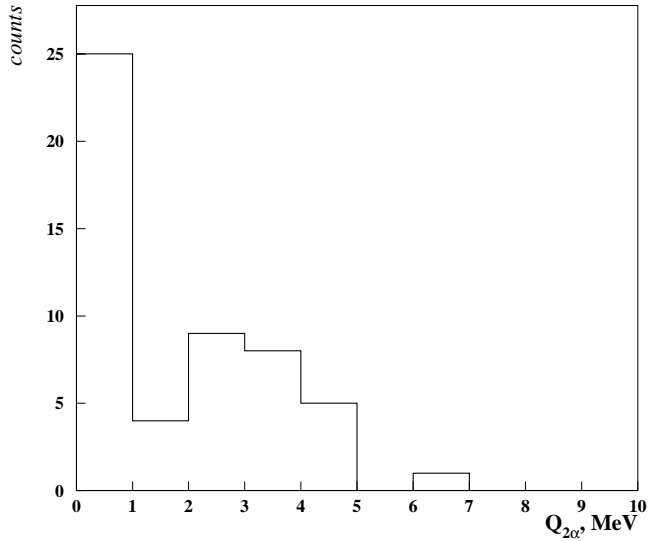}
}
\caption{Distribution of $\alpha$ particle pairs $vs$ $Q_{2 \alpha}$ (3) for the fragmentation mode
 $^9$Be$\rightarrow$2$\alpha$.}
\label{fig:7}
\end{figure}
\begin{figure}
\resizebox{0.5\textwidth}{!}
{
  \includegraphics{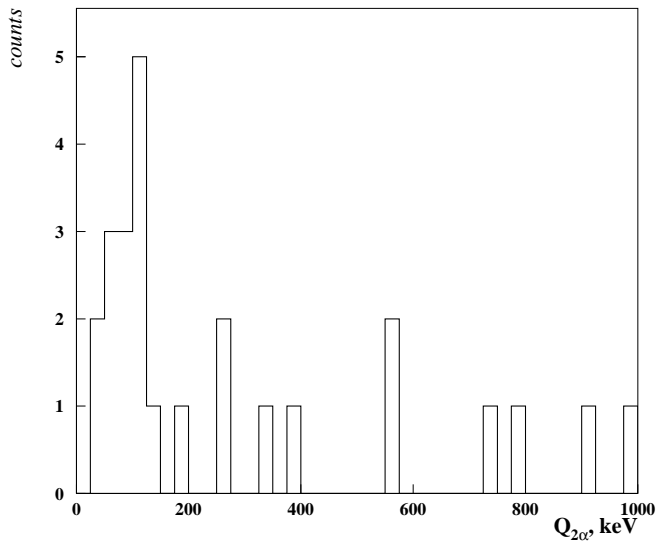}
}
\caption{Distribution of $\alpha$ particle pairs $vs$ Q$_{2 \alpha}$ (3) for the fragmentation mode $^9$Be$\rightarrow$2$\alpha$ 
zoomed between 0-1000~keV.}
\label{fig:8}
\end{figure}

\indent   
Fig.~\ref{fig:7} shows the $Q_{2 \alpha}$ distribution  (3) allowing one to estimate the excitation scale. There is an
 event concentration below 1~MeV which is relevant for the $^8$Be ground
 state decay. Besides, one can resolve a bump at around 3~MeV corresponding to the $^8$Be decay from the
 first excited state 2$^{+}$.
 A zoomed part of this distribution near zero is presented in fig.~\ref{fig:8}. A clear peak is seen as a concentration of
 14 events around  the mean value $Q_{2 \alpha}$ equal to 88 keV which is close to the decay energy of the $^8$Be
 ground state. Thus, the achieved identification of $^8$Be production allows one to justify
 the spectroscopy of $n\alpha$ decays from the lowest decay energy.\par
\vspace{3.5cm}
\section{\label{sec:level6}Charged topology of $\bf^8$B fragmentation}
\indent 
In order to expose an emulsion to $^8$B nuclei, the use was made of fragmentation of
 1.2~A~GeV $^{10}$B nuclei. In this case, absence of $^9$B nuclei among secondary fragments turned out to be good luck, resulting in a clear
 separation of the primary and secondary
 beams  by their magnetic rigidity. When scanning emulsions, this fact was indirecly confirmed by the
 absence of \lq\lq white\rq\rq ~stars with a charge topology He+H. They
 could be produced by background $^6$Li having the same magnetic rigidity as $^{10}$B nuclei. A 15\% -admixture of $^7$Be nuclei
 was eliminated according to the charge topology of the found \lq\lq white\rq\rq ~stars. The most intensive background 
 presented by  $^3$He nuclei was visually separated.\par

\begin{table}
\caption{\label{tab:2}Charge-topology distribution of the \lq\lq white\rq\rq ~stars originated from the dissociation 
of 2.1~A~GeV $^{8}$B nuclei.}
\begin{tabular}{l|c|c|c|c|c}
\hline\noalign{\smallskip}
\hline\noalign{\smallskip}
$Z_f$    & 4  & 3  & -- & -- & -- \\
$N_{2}$  & -- & -- & 2  & 1  & -- \\
$N_{1}$  & 1  & 2  & 1  & 3  & 5  \\
$N_{ws}$ & 15 & 1  & 9  & 11 & 3  \\
\hline\noalign{\smallskip}
\hline\noalign{\smallskip}
\end{tabular}
\end{table}
\indent By scanning over the incoming tracks, a total of 39 \lq\lq white\rq\rq ~stars, in which the 
 charge in the 15$^{\circ}$-cone is equal to 5, have been found. 
 Their distribution by the charge modes is shown in table~\ref{tab:2} in the same manner as in table~\ref{tab:1}.
 The significance of the  $^8$B modes can be compared with 
 the  topology of \lq\lq white\rq\rq ~stars produced by the 1~A~GeV $^{10}$B nuclei  \cite{Adamovich04}. 
The fraction of the 
 3-prong stars $^{10}$B$\rightarrow$ 2He+$^{1,2}$H was established to be equal to 80 \% with 40\% deuteron clustering.
The probability of the 2-prong mode $^{10}$B$\rightarrow$ $^{9}$Be+$^{1}$H was
 found to be equal only to 3\%. Such a strong difference can be explained
 by the lower value of the $^{2}$H binding energy.\par
 \indent As is shown in table 2, an obvious distinction of the $^{8}$B  case consists in a high yield of the 
2-prong mode $^{8}$B$\rightarrow$ $^7$Be+$^{1}$H. This feature is due to the weak
 $^{1}$H binding.  Thus, one can conclude that the loosely bound $^{8}$B nucleus 
manifests its structure already in the charge-topology.
Further it is planned
 to increase statistics, identify the H and He isotopes, and
 reconstruct emission angles.\par
 \indent Obtaining $^{10,11}$B beams at the JINR Nuclotron makes it possible to form
 $^{10,11}$C secondary beams by the use of charge-exchange reactions analogous to the 
 $^7$Li$\rightarrow$ $^7$Be process \cite{Bradnova04}.
 This method is optimal for the 
 emulsion technique where of importance
 is the simplicity of identification of incoming nuclei 
rather than their intensity. 
 The existence and the cross sections of such
 processes will be established in a separate experiment.
 The suggested emulsion exposures will allow one to explore the 3-prong modes $^{10,11}$C$\rightarrow$3He analogous to the case
 $^{12}$C$\rightarrow$3$\alpha$ \cite{Belaga95}. Clustering in $^{12}$C$\rightarrow$3$\alpha$ reflects the ternary $\alpha$ processԮ 
Study of the 3He clustering in $^{10,11}$C fragmentation would serve as a basis  for studying the possible role of the ӳHeԠfusion process  
in nucleosynthesis, that is, in media with a mixed composition 
of He isotopes. 
\par 
\section{\label{sec:level7}Charged topology of $\bf ^9$C fragmentation}
\indent Unfortunately, it is impossible to use the approach based on a charge-exchange reaction for the formation of
 a $^9$C nucleus beam. An emulsion was exposed to a
 secondary beam produced in fragmentation of the 2.1~A~GeV $^{12}$C nuclei and having a $^9$C magnetic
 rigidity. A search was made for the events in
 which the total charge of tracks in a forward 15$^{\circ}$ cone is equal to 6. The presently found 17 events
 are distributed as shown in table~\ref{tab:3}. The 3He mode, which dominantes the $^{12}$C \lq\lq white\rq\rq ~stars,
 is seen to be suppressed. Besides, there is an indication that multiple dissociation channels predominate over
possible candidates to the p+$^{8}$B and p+p+$^{7}$Be modes. Verification of this observation makes a farther 
increase in statistics to be particularly intriguing.\par
\indent The cases, which might be interpreted as 3$^3$He, are of especial importance since they point to highly-lying
 cluster excitations associated with a
 strong nucleon rearrangement to produce the 3$^3$He system. Like the process $^{12}$C$\rightarrow$3$\alpha$, this
 dissociation can be considered as
 a visible reflection of the inverse process of a ternary $N_{ws}$He fusion. It can provide a
 significantly higher energy output
 followed by $^4$He pair production. Search for an ternary $^3$He process appears to be a major goal of
 further accumulation of statistics.\par
 \begin{table}
 \caption{\label{tab:3}Charge-topology distribution of the \lq\lq white\rq\rq ~stars originated from the dissociation 
of 2.1~A~GeV $^{9}$C nuclei.}
\begin{tabular}{l|c|c|c|c|c}
\hline\noalign{\smallskip}
\hline\noalign{\smallskip}
$Z_f$    & 5  & 4  & -- & -- & --  \\
$N_{1}$  & 1  & 2  & -- & 2  & 4  \\
$N_{2}$  & -- & -- & 3  & 2  & 1   \\
$N_{ws}$ & 1  & 2  & 3  & 7  & 5   \\
\hline\noalign{\smallskip}
\hline\noalign{\smallskip}
\end{tabular}
\end{table}
\section{\label{sec:level8}Conclusions}
\indent The invariant approach is applied to analyzing the relativistic
 fragmentation of $^{22}$Ne, $^{14}$N and $^9$Be nuclei having a significant difference in the primary energy.
 It is shown that doubly charged fragments having relative $b_{ik}$ within the range 
$b_{ik}<$10$^{-2}$ form well-separated $n\alpha$ jets.
 It corresponds to the relative motion of $\alpha$ particles with relative kinetic energy
 of the order of 1 MeV per nucleon in the jet center-of-mass system. \par
\indent New experimental observations are reported from the emulsion exposures to
 $^{14}$N and $^9$Be nuclei with energy above 1~A~GeV. Being applied to the
 fragmentation of these nuclei the invariant analysis is shown to be a promising means to study excited states of simple  $\alpha$ particle systems.
 The internal energy of a system involving He fragments can be estimated in an invariant form
 down to the $^8$Be nucleus decays.\par
\indent  The pattern of the relativistic fragmentation becomes
 more complete in the case of proton excess in the explored nucleus. It is shown that 
nuclear track emulsions provide unique possibilities to
explore few-body decays of  $^{8}$B and $^{9,10,11}$C nuclei.
 The paper describes the start of this work. The invariant approach applied for the stable nuclei
 will be of special benefit in the case of 
the neutron deficient nuclei.\par   
\indent In spite of statistical restrictions, nuclear track emulsions ensure the initial stage
 of investigations in an unbiased way and enable
 one to develop scenarios for dedicated experiments \cite{web}.
 Our experimental observations concerning few-body aspects of nuclear physics can be described in the
 relativistic invariant form allowing one to
 enlarge nuclear physics grounds of the nucleosynthesis pattern.\par
 \indent The work was supported by the Russian Foundation for Basic Research (Grants nos. 96-1596423,
 02-02-164-12a, 03-02-16134, 03-02-17079, 04-02-16593, 04-02-17151),
 the Agency for Science of the Ministry for Education of the Slovak Republic and the Slovak Academy
 of Sciences (Grants VEGA 1/9036/02 and 1/2007/05) and  grants from the JINR Plenipotentiaries
 of Bulgaria, Czech Republic, Slovak Republic, and Romania  during 2002-5.\par

\end{document}